# Emergent Information Formation in Prebiotic Protocell Clusters:
# A Computational Mechanics Framework of ε-Machines and Attractor Memory

Michael Massoth
Department of Computer Science, Hochschule Darmstadt (h_da)
University of Applied Sciences Darmstadt, member of European University of Technology (EUt+)
Darmstadt, Germany
e-mail: michael.massoth@h-da.de

*Abstract*- **Casimir–Lifshitz forces generate an unavoidable, long-range attraction between protocells under prebiotically realistic conditions. This interaction stabilizes mesoscale clusters such as tetrahedra, octahedra, and 13-cell icosahedra. These highly symmetric assemblies act as persistent macrostates whose transitions remain reproducible despite microscopic noise. A physics-guided coarse-graining yields a well-defined mesodynamics that can be represented as an ε-machine: a small deterministic automaton whose causal states correspond to cluster attractors and whose transitions encode ordered reconfiguration pathways. The theory of Rosas et al. ("Software in the natural world") shows that such systems can become informationally, causally, and computationally closed, thereby forming an autonomous proto-software layer. In this framework, prebiotic information does not arise from polymers but from attractor-based memory and structured transition dynamics in a purely physical cluster process.**

## I. INTRODUCTION

This is the second of seven papers in the series: "A Constructivist Proto-Bio-Information Theory: A Physically Grounded Nano-Systems Architecture for Prebiotic Emergence, Information, Proto-Semantic Function, and Sustainability of Protocell Aggregation and Cluster Formation".

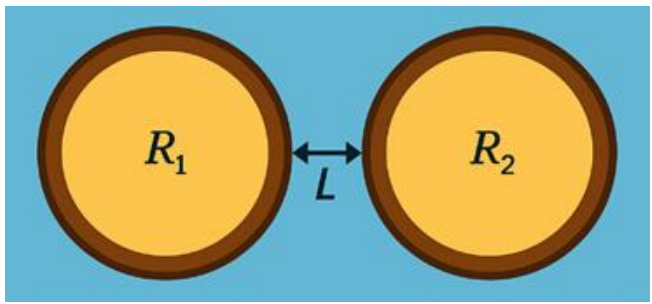

Figure 1. *Schematic of two prebiotic protocells with radii $R_1$ and $R_2$ and minimal surface-to-surface separation L in saline water (primordial soup).*

In Figure 1, a schematic representation of two spherical protocells with radii $R_1$ and $R_2$ is shown, separated by a minimal surface-to-surface distance L in saline aqueous solution (primordial soup). This sphere–sphere geometry defines the fundamental configuration used to model Casimir–Lifshitz interactions between protocell membranes under prebiotic conditions.

Casimir–Lifshitz forces arise from quantum and thermal fluctuations of the electromagnetic field between material interfaces, as first described by Hendrik Casimir [3] and generalized by Evgeny Lifshitz [4]. Unlike chemical bonds or electrostatic interactions, they do not rely on charges, receptors, or molecular specificity, but on the dielectric properties of the interacting materials and medium. In saline, thermally active environments, these forces act over nanometre-to-submicrometre distances and decay algebraically rather than exponentially. They therefore provide a physically unavoidable, environment-robust interaction mechanism precisely in the mesoscale regime where classical colloidal forces fail.

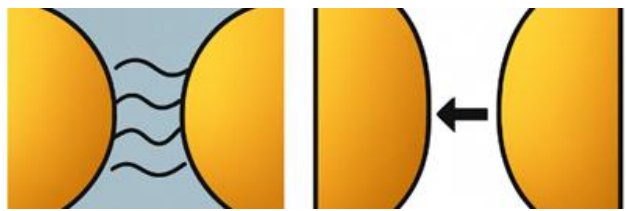

Figure 2. Thermal and quantum field fluctuations (left) generate Casimir-Lifshitz forces (right), which stabilize neighboring protocell membranes at nanoscale distances (2–100 nm) as clusters.

In Figure 2, quantum and thermal electromagnetic field fluctuations (left) give rise to Casimir–Lifshitz forces (right) between neighboring protocell membranes. These fluctuation-induced interactions act over nanometre-to-submicrometre separations ($\approx$ 2–100 nm) in saline environments and provide a robust, non-chemical mechanism for stabilizing mesoscale protocell clusters where classical colloidal forces fail.

Massoth [1] showed that attractive Casimir–Lifshitz forces represent a universal and unavoidable aggregation mechanism under prebiotically realistic conditions. For protocell radii of R = 200–1000 nm, separations L = 5–100 nm, salt concentrations of 50–200 mMol, and temperatures of 20–90

°C, the classical DLVO electrostatic repulsion is almost fully suppressed. This is due to the short Debye length (< 2 nm) in saline environments. Van der Waals interactions act only at 0.3–3 nm and therefore cannot stabilize larger mesoscale assemblies. In contrast, Casimir‑Lifshitz forces (Figure 3) retain an algebraic range of approximately ~$1/L^2$ and remain effective even in highly ionic and thermally agitated media. The strong dielectric contrast between protocell membranes ($\varepsilon \approx 2\text{–}8$) and water ($\varepsilon \approx 75$) produces a robust, attractive fluctuation-induced force that persists throughout the prebiotic parameter space.

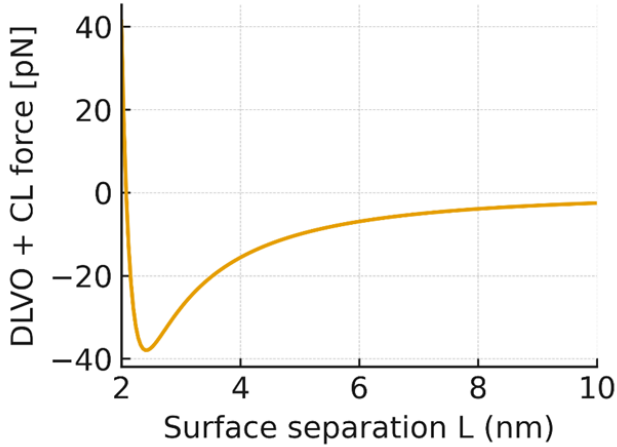

Figure 3. Illustrates the resulting net interaction force in the range of *2–10 nm*. The algebraic Casimir–Lifshitz contribution creates a residual, non-DLVO attraction that remains detectable experimentally.

The corresponding potential wells $U_{CL}(L)$ are sufficiently deep to stabilize dimers, trimers, and higher-order clusters. Larger protocells couple more strongly because of their increased effective curvature radius $R_{eff}$. Membrane types with higher polarizability—such as PMBCs—generate deeper potential wells than simple fatty-acid vesicles and should therefore cluster more readily. Expected mesoscale contact times fall in the minute range, and a temperature window of 40–80 °C maximizes stability.

These results indicate that attractive Casimir–Lifshitz forces provided a purely physical mode of early cooperation long before genetic or metabolic mechanisms existed. Protocell clusters may thus have acted as proto-ecological units from which functional and information-bearing structures later emerged.

Structure of the paper:
In Section 2, we derive Casimir–Lifshitz interaction scaling for spherical protocells and quantify cluster stabilization energies across "magic-number" geometries.
In Section 3, we position this mechanism within prior work on prebiotic self-organization, semantic information, and computational mechanics.
In Section 4, we introduce a coarse-grained description $Z_t=f(X_t)$ and show how Casimir-stabilized cluster attractors define recurrent macrostates with enhanced predictability.
In Section 5, we apply the Rosas framework [2] to establish informational, causal, and computational closure and identify the corresponding ε\varepsilonε-machine structure.
In Section 6, we formalize the micro-to-macro reduction from $X_t$ and connect dimensionality reduction to conditional-entropy compression.
In Section 7, we analyze attractor-based memory and retrieval dynamics in protocell clusters and relate return paths to robust mesoscale storage.
In Section 8, we construct the explicit ε-machine representation of cluster-state dynamics and its transition structure.
In Section 9, we synthesize these results into a physically grounded account of prebiotic information prior to genes or polymer-based coding.
In Section 10, we unify the closure concepts and ε-machines into a single multilevel framework for prebiotic emergence.
In Section 11, we summarize implications and outline testable predictions and future theoretical and experimental directions.

## II. MESOSCALE PROTOCELL CLUSTERS STABILIZED BY CASIMIR–LIFSHITZ FORCES

For protocells with radii $R \approx 200\text{–}1000$ nm, Casimir–Lifshitz attraction at separations of a few nanometers drives robust self-assembly into densely packed mesoscale clusters. For two spherical protocells (a dimer) with radii $R_1$ and $R_2$ and minimal surface separation *L*, the effective potential is: $U_{CL}(L) \approx -(A_{eff}/6) * (R_{eff}/L)$, with $A_{eff}$ the effective Hamaker constant of the membrane-water-membrane system and $R_{eff} = (R_1 * R_2)/(R_1 + R_2)$. For equal-sized protocells ($R_1 = R_2 = R$) this reduces to: $U_{CL}(L) \approx -(A_{eff}/12) * (R/L)$.

The natural scale is given by the dimensionless coupling strength: $\Lambda(R,L) = |U_{CL}(L)| / k_BT = (A_{eff} * R) / (12 k_BT * L)$. For realistic values ($A_{eff} \approx 5\times10^{-21}$ J, $R \approx 500$ nm, $L = 3\text{–}30$ nm), one obtains $\Lambda \approx 1\text{–}10$. The Casimir–Lifshitz attraction therefore exceeds thermal energy in the relevant distance range and can effectively suppress Brownian fluctuations.

The total energy of an *N*-particle cluster in near-contact follows from the number of particle–particle bonds $N_{bonds}(N)$: $E_N(L) \approx -N_{bonds}(N) * |U_{CL}(L)| = -N_{bonds}(N) * (A_{eff}/6) * (R_{eff}/L) = -N_{bonds}(N) * \Lambda(R,L) * k_BT$.

TABLE I. TOTAL CASIMIR–LIFSHITZ INTERACTION ENERGIES

| N | Protocell Structure | $N_{bonds}$ | $E_N/k_BT \approx$ |
|---|---|---|---|
| 2 | Dimer | 1 | −5.1 |
| 3 | Triangularer Trimer | 3 | −15.2 |
| 4 | Tetrahedron | 6 | −30.5 |
| 6 | Oktahedron | 12 | −61.0 |
| 7 | Pentagonal Bipyramid | 15 | −76.2 |
| 13 | Icosahedral 13-Cluster | 42 | −213.4 |

Table I, shows the total Casimir–Lifshitz interaction energies $E_N/k_BT$ for representative N-protocell cluster geometries,

illustrating how the number of pairwise bonds Nbonds drives mesoscale stabilization.

For isotropic, short-range attraction, the system favors "magic-number" clusters that maximize contacts: Dimer (N = 2, $N_{bonds}$ = 1), Triangular Trimer (N = 3, 3 bonds), Tetrahedron (N = 4, 6 bonds), Octahedron (N = 6, 12 bonds), Pentagonal bipyramid (N = 7, 15 bonds). On the next mesoscale level, a 13-protocell icosahedral cluster (12 around 1 center) forms with $N_{bonds}$ = 42. This structure is exceptionally symmetric and tightly packed.

Using *R = 500 nm, L = 10 nm* and $A_{eff}$ = *5×10$^{-21}$ Joule*, cluster binding energies range from approximately $-15.2\ k_BT$ (triangular trimer) to more than $-200\ k_BT$ (icosahedral 13-cluster). These values show that Casimir–Lifshitz attraction is strong enough not only to stabilize dimers and oligomers but to enforce the formation of robust mesoscale clusters that can serve as seeds for further hierarchical structure formation. Experimental studies of primitive vesicle compartments further support the plausibility of such prebiotic mesoscale assemblies [13].

### III. RELATED WORK

Classical approaches to prebiotic organization emphasize chemical self-organization or polymer-based information, while physical interactions between early compartments are often treated as transient aggregation effects. In parallel, Computational Mechanics (J. P. Crutchfield & C. R. Shalizi, *"Computational Mechanics: Pattern and Prediction, Structure and Simplicity"*) [8][11] formalizes syntactic information via ε-machines, yet leaves the physical origin of causal states unspecified.

Recent theoretical approaches to semantic information—most notably the work of Kolchinsky and Wolpert (*"Semantic Information, Autonomous Agency and Non-Equilibrium Statistical Physics"*, 2018) [16] and of Ruzzante et al. (*"Synthetic Cells Extract Semantic Information From Their Environment"*, 2023) [17][18] - demonstrate that information is not intrinsically tied to genes or replication. However, they leave open which concrete physical structures carry semantically relevant states. The models operate predominantly at an abstract agent–environment level or focus on individual synthetic cells.

This paper advances these frameworks by identifying Casimir–Lifshitz–stabilized protocell clusters as a concrete physical substrate for syntactic information. Highly symmetric mesoscale clusters form robust attractors whose transitions remain reproducible despite microscopic noise. A physics-guided coarse-graining reconstructs these attractors as ε-machine states, yielding informational closure in the sense of Rosas et al., "Software in the Natural World" (2024) [2]. Unlike abstract models, information here arises directly from physical self-organization, without genes, polymers, or symbolic encoding.

### IV. EMERGENCE OF MESOSCALE INFORMATION IN CASIMIR–LIFSHITZ–STABILIZED PROTOCELL CLUSTERS

Early prebiotic compartments lacked genes, enzymes, and regulatory networks. Their organization depended entirely on physical processes. Yet some structures must already have been stable over time, recurrent, and minimally coherent. A natural candidate for such early organization is the set of mesoscale cluster states formed by aggregating protocells.

We now apply the computational-mechanics framework of ε-machines and attractor memory [2] to formalize how Casimir–Lifshitz–driven protocell coupling gives rise to emergent informational states.

Casimir–Lifshitz forces stabilize a small number of highly symmetric, recurrent configurations on the mesoscale. Typical examples include dimers, trimers, tetrahedra, octahedra, and 13-protocell icosahedra. These assemblies act as discrete macro-states $Z_t$ that reappear reliably despite microscopic noise and persist over mesoscale timescales. Contact times are typically on the order of minutes, making these geometries robust attractors in configuration space.

These macro-states are not static objects. Clusters can rearrange and transition between configurations, yet remain confined to a small, stable state space. This yields a dynamic but structured mesodynamics, in which only a limited set of configurations dominates.

This situation naturally suggests a coarse-graining in which the high-dimensional microdynamics Xt—positions, forces, membrane fluctuations, and hydrodynamic interactions—collapse onto a finite set of functional states: $Z_t$ = f($X_t$). The identification of such predictive macro-states directly corresponds to the construction of causal states in the sense of Crutchfield's statistical complexity framework [8]. A key consequence of this reduction is a strong increase in predictability. Even though the micro-dynamics are noisy, the meso-dynamics follow clear patterns. The system's future is far better predicted from $Z_t$ than from the full microstate. Formally, this appears as a marked decrease in conditional entropy: $H(Z_{t+1} \mid Z_t) \ll H(Z_{t+1} \mid X_t)$, which implies informational closure. The macro-level acquires its own lawful dynamics, largely independent of microscopic details.

Within Computational Mechanics, this organized behavior is represented as an ε-machine. Its causal states $E_i$ group all histories that yield identical future distributions. Each causal state corresponds to a meaningful cluster geometry or transition motif. Transitions $E_i \rightarrow E_j$ represent allowed transformations, such as the relaxation of a distorted octahedron or the completion of a metastable 11-cluster into a 13-icosahedral configuration.

A defining feature of this dynamics is its attractor-based memory. Cluster geometries act as robust storage states that the system reliably returns to after perturbations. These deterministic return paths match the "memory retrieval dynamics" described by Rosas et al.: the spatial configuration itself encodes past trajectories and functions as a physical information unit without genes or polymers.

This yields a physically grounded and mathematically consistent model of prebiotic information formation. Protocell clusters exhibit a software-like meso-dynamics characterized by discrete macro-states, stable attractors, and an ε-machine structure. Prebiotic information arises not from symbolic coding but from mesoscale, dissipative, attractor-based structural persistence—a necessary precursor to later molecular complexity.

## V. APPLICATION OF THE COMPUTATIONAL MECHANICS FRAMEWORK: EMERGENCE, CLOSURE, AND ε-MACHINES IN PROTOCELL CLUSTERS

The computational mechanics framework of Rosas et al. [2] provides a precise method for quantifying emergence in natural systems. It distinguishes three levels of autonomous organization—informational, causal, and computational closure—which capture how a macro-level becomes independent of microscopic details. This structure is well suited for prebiotic protocell clusters, whose Casimir–Lifshitz–stabilized attractors generate a coherent, software-like meso-dynamics.

*Informational closure* occurs when the future of a macro-state $Z_t$ is better predicted from $Z_t$ itself than from the full micro-dynamics $X_t$. In protocell clusters, transitions between stable geometries—such as tetrahedron → octahedron or an 11-cluster → 13-icosahedron—are largely insensitive to local fluctuations. Formally: $H(Z_{t+1} | Z_t) \ll H(Z_{t+1} | X_t)$.

Even under thermal noise, the system reliably returns to the same attractors. This marks the first level of emergent autonomy.

*Causal closure* requires that the macro-level causes its own future. This notion aligns with the theory of causal emergence, which shows that macro-descriptions can exert greater effective causal influence than their underlying microstates [9]. The global cluster structure—geometry, contact graph, coupling strengths—governs the dynamics, not the precise microstate. This is reflected in the equivalence of micro- and macro-interventions:

$P(Z_{t+1} | do(Z_t)) = P(Z_{t+1} | do(X_t \in pre(Z_t)))$

where $pre(Z_t)$ is the set of microstates compatible with the same cluster pattern. Thus, the macroprocess exhibits genuine causal power, not merely statistical regularity.

*Computational closure* denotes the ability of the macro-level to form its own internal representation. In Computational Mechanics, this is the *ε-machine*: a minimal deterministic automaton whose causal states $E_i$ collect all pasts with identical future distributions. For protocell clusters, these $E_i$ correspond to stable and metastable geometries—dimer, trimer, tetrahedron, octahedron, 13-icosahedron—and their transitions. The transition structure $P(E_i \rightarrow E_j)$ defines a small finite-state graph capturing the cluster's complete dynamic rules.

A core result of the Rosas framework [2] is that informational closure automatically implies computational closure. Once the future can be reliably inferred from $Z_t$, a consistent ε-machine emerges. For the prebiotic world, this means that the physical attractors of protocell clusters form an emergent rule-level analogous to an early "software layer." Clusters store history in their geometry, follow reproducible transition paths, and possess a robust dynamical memory without genes or polymers.

Overall, Casimir-stabilized protocell clusters generate not only stable mesoscale structures, but also an autonomous macro-level with its own causality and internal representation. Emergence, closure, and ε-machines together offer a coherent theoretical basis for a physically grounded model of prebiotic information—long before genetic coding, enzymatic control, or molecular replication existed.

## VI. COARSE-GRAINING PROTOCELL DYNAMICS: FROM MICROSTATES $X_T$ TO MACROSTATES $Z_T$

The microphysical dynamics of prebiotic protocells are high-dimensional. Positions, radii, membrane fluctuations, Casimir–Lifshitz forces, hydrodynamic couplings, and thermal noise create a deterministic–stochastic system with a large phase space. A microstate $X_t$ includes spatial coordinates, force fields, local membrane structure, and solvent fluctuations - a level of detail that obscures emergent order.

A biologically meaningful coarse-graining reduces this complexity by grouping microstates into functional macro-states $Z_t$. These macro-states describe protocells not individually, but as reconfiguring clusters with typical sizes N={2,3,4,6,7,13,…}, characteristic geometries (tetrahedral, octahedral, icosahedral), coupling strengths Λ(R,L), contact persistence's, and stable neighborhood topology. Such variables are insensitive to microscopic fluctuations and remain coherent over mesoscale times.

Formally, the reduction is given by $Z_t = f(X_t)$, where distinct microstates $X_t$ and $X_{t'}$ map to the same macro-state and yield nearly identical future behavior. The macro-states thus take on a *semantic* role: they are the dynamically relevant units of cluster organization.

This coarse-graining reduces dimensionality dramatically. The emergence of low-dimensional macro-states from high-dimensional microscopic physics is a hallmark of dissipative structure formation in non-equilibrium systems [12]. Although $X_t$ is continuous and high-dimensional, the Casimir–Lifshitz energy landscape supports only a small set of stable $Z_t$. Dimers, tetrahedra, and 13-icosahedra act as attractors of the state space. An attractor A satisfies: $P(Z_{t+1} \in A | Z_t \notin A) \rightarrow$ high, and $P(Z_{t+1} \in A | Z_t \in A) \rightarrow$ very high.

These attractors correspond to real physical configurations. They minimize energy, absorb fluctuations, and restrict the macro-state space to a small, recurring set with robust transitions.

From an information-theoretic perspective, macroscopic dynamics occupy a reduced, lower-dimensional state space and thus constitute a compressed representation of the underlying microscopic fluctuations, exhibiting substantially higher predictability than the far more complex micro-dynamics. This follows from the reduced conditional entropy:

$H(Z_{t+1} | Z_t) < H(X_{t+1} | X_t)$,

indicating a structured, attractor-based dynamics similar to neural systems and other dissipative self-organizing media.

The result is a set of robust, repeatable, and causally effective macro-states - the foundation of a proto-informational dynamics in which structure and function are physically coupled without genes or molecular codes. The macro-level of protocell clusters therefore represents an early form of physical organization: a prebiological "software layer" emerging from energetic and geometric self-organization, preparing the ground for later evolutionary complexity.

## VII. MEMORY RETRIEVAL AND ATTRACTOR DYNAMICS IN PROTOCELL CLUSTERS

A key result of Rosas et al. [2] is the presence of memory in systems with multiple stable attractors. Case Study F shows that noise-driven systems fall into energetic minima and form dynamic memory: past states bias transitions, positions in attractor space define functional states, and return paths after perturbations are reproducible. This "memory-retrieval" behavior appears when the macro-level is resilient to microscopic noise.

Casimir–Lifshitz-stabilized protocell clusters show the same pattern. Their fluctuation forces create a discrete set of energetic attractors—dimers, tetrahedra, octahedra, and the deep 13-icosahedral minimum (below $-200$ $k_B T$). These geometries act as local minima and remain highly likely under prebiotic conditions. This behavior reflects classical principles of self-organization and attractor landscapes in far-from-equilibrium systems, as described by Kauffman [10].

The transition from a microstate $X_t$ into an attractor can be described as motion within a basin-of-attraction structure. Let $A_i$ denote the set of microstates leading to attractor. Under prebiotic parameters: $P(Z_{t+1} = i | X_t \in A_i) \approx 1$.

Small deformations from hydrodynamic or thermal noise drive the cluster along characteristic return paths back into the same attractor. Such energy-minimizing return dynamics closely parallel attractor-based memory retrieval in Hopfield networks [15]. These relaxation trajectories correspond directly to the "retrieval trajectories" of Rosas et al. [2].

Memory dynamics have two components. First, the cluster geometry acts as a physical storage medium: a distorted octahedron relaxes back to its native form, minimizing the conditional entropy along the return path,

$H(Z_{t+1} | Z_t \in \text{Attractor}_i) \to$ minimal.

Repeated transitions also reinforce future behavior. Frequent returns to the same attractor reshape the ε-machine's transition matrix and increase recurrence. Attractors become stable patterns that store "proto-historical information" across many fluctuation cycles.

This memory needs no replication, templates, or polymers. The information is not symbolic; it is encoded in the cluster's geometry and topology. Because transitions are constrained and macrostates persist, the ε-machine forms a proto-software layer: attractor states act as nodes, retrieval paths as directed edges.

The macro-level behaves like a physical memory system. The cluster "knows" its attractor and reliably returns after perturbations. Past shapes and relaxation paths bias future transitions. This mutual dependence is characteristic of attractor-based memory and the proto-software systems described by Rosas et al. [2].

Thus, Casimir–Lifshitz-stabilized protocell clusters show a definable memory mechanism. Attractors serve as storage states, return trajectories as retrieval functions. Prebiotic matter therefore possessed a purely physical principle of information and memory — a software-like layer without genes or molecules, forming a foundation for later biological evolution.

## VIII. PROTOCELL CLUSTERS AS AN ε-MACHINE

The emergent dynamics of prebiotic protocell clusters can be described not only qualitatively through attractors but also quantitatively using Computational Mechanics. This framework models the mesoscale cluster states and their transitions as an ε-machine — a minimal deterministic automaton that captures the full causal structure of a process. Casimir–Lifshitz-stabilized protocells provide ideal conditions: they form a small, discrete set of long-lived macro-states whose transitions remain largely independent of microscopic variations. This yields a clear, formally defined notion of "proto-software" arising purely from physical self-organization.

The system is first represented as a sequence of macro-states $Z_t$. These include stable cluster geometries — dimer, trimer, tetrahedron, octahedron, 13-icosahedron — and metastable transitions such as partial recombinations or fragmentations:

$Z_0, Z_1, Z_2, \ldots, Z_t \in$ {dimer, trimer, tetrahedron, octahedron, …, decay, recombination, …}

Each macro-state summarizes many microconfigurations $X_t$ that are functionally equivalent. Although the micro-dynamics is high-dimensional, the future is far more predictable from $Z_t$ than from $X_t$ — a key signature of macro-level stability.

Next, causal states $E_i$ of the ε-machine are constructed. This definition follows the formalism of Shalizi & Crutchfield, who defined ε-machines as the minimal sufficient representation of a process [11]. Two past trajectories $z(t|-\infty:0)$ and $z'(t|-\infty:0)$ are causally equivalent when they generate identical future distributions: $P(Z_{t+1} | z(t|-\infty:0)) = P(Z_{t+1} | z'(t|-\infty:0))$.

Transitions between causal states are given by $T_{ij} = P(E_j | E_i)$.

Because cluster attractors are energetically stable, the automaton contains only few states and exhibits characteristic attractor cycles — a hallmark of emergent software-like dynamics in the sense of Rosas et al.

Testing closure properties confirms that the ε-machine defines an autonomous macro-level:

*Informational closure:* $H(Z_{t+1} | Z_t) \ll H(Z_{t+1} | X_t)$, showing that the macrostate predicts the future better than the microstate.

*Causal closure:* $P(Z_{t+1} | do(Z_t)) = P(Z_{t+1} | do(X_t \in pre(Z_t)))$, demonstrating that macro-interventions and equivalent micro-interventions yield the same transitions.

*Computational closure:* $\varepsilon(f(X_t)) = \pi(\varepsilon(X_t))$, where f is the macro-mapping and π the projection onto causal classes. This shows that coarse-graining produces the same causal architecture as full microanalysis.

Thus, the cluster dynamics establishes an autonomous computational layer — a proto-software system not based on symbolic coding but on stable attractors and reproducible transitions. In this strict formal sense, the ε-machine represents prebiotic information: a closed set of causal rules arising entirely from physical self-organization in mesoscale protocell ensembles, forming the foundation of later biological information processing.

## IX. Physically Grounded Origin of Prebiotic Information

The origin of information is a central question in origins-of-life research. Compartment-based models of early cellular organization emphasize that structural stability itself may precede genetic information [14]. Classical models tie information to polymers, such as RNA or DNA. Yet before such molecules existed, systems still needed structures that were stable, recurrent, and capable of minimal functional persistence. The physical dynamics of protocell clusters offer a precise starting point.

Information emerges whenever a system forms stable, distinguishable, and reproducible macro-states whose transitions follow their own dynamics. A macro-state $Z_t$ is informative when it predicts the future better than the underlying microstate $X_t$: $H(Z_{t+1} | Z_t) < H(Z_{t+1} | X_t)$.

Casimir–Lifshitz-stabilized protocell clusters satisfy this condition strongly. Their attractors — tetrahedra, octahedra, 13-icosahedra — are stable over many $k_B T$, recur frequently, and are largely insensitive to microscopic fluctuations. Transitions follow ordered paths: tetrahedra often relax into octahedra, incomplete 11-clusters complete into 13-icosahedra, and decay trajectories follow consistent routes.

These attractors also store memory. A slightly deformed octahedron relaxes deterministically back into the same attractor. Formally, this appears as a contraction in attractor space: $\|Z_{t+1} - A\| < \|Z_t - A\|$.

Thus, attractors function as both stable and memory-bearing states — an *attractor-based memory*, in the sense of Rosas et al. [2]. The ε-machine makes this structure explicit. The macro-states $Z_t$ form a finite set of causal states $E_i$, and the transition probabilities $T_{ij}$ define an internal, proto-symbolic rule structure. No polymers are required. Instead, information is stored in the geometry, topology, and allowed transitions of the cluster.

This yields a prebiotic, non-gentic form of information arising purely from physical self-organization. Protocell clusters behave as physical information carriers whose attractor-based transition dynamics generate an early "software layer" — an information-processing mechanism without genes or enzymes, yet with clear functional structure and evolutionary potential.

## X. ε-Machines and Closure Concepts as a Unified Framework for Prebiotic Emergence

The combination of Casimir–Lifshitz stabilization, mesoscale cluster attractors, and the theory of Computational Mechanics provides a coherent model for the emergence of prebiotic information. The framework links micro-, meso-, and macro-levels and shows how protocell clusters can develop a software-like dynamics generated purely by physical self-organization. Central to this model are the three closure concepts of Rosas et al.—informational, causal, and computational closure—together with the ε-machine formalism. The resulting hierarchy is:

TABLE II. MULTI-LEVEL HIERARCHY OF PREBIOTIC PROTOCELL DYNAMICS

| Level | Description | Formalization |
|---|---|---|
| Micro | Forces, positions, thermal noise | Stochastic dynamics $X_t$ |
| Meso | Cluster attractors | Macroprocess: $Z_t = f(X_t)$ |
| Macro | Information dynamics | ε-machine $E_t$, transitions $T_{ij}$ |
| Software layer | Attractor-coded memory, pattern retrieval | Causal / computational closure |

Table II shows the hierarchical organization of protocell dynamics from microphysical interactions through mesoscale cluster attractors to an emergent macro-level ε-machine and a proto-informational software layer characterized by causal and computational closure.

On the micro level, Casimir–Lifshitz forces, hydrodynamic coupling, and thermal noise shape protocell motion. The dynamics $X_t$ is high-dimensional and difficult to predict.

A biologically motivated coarse-graining maps many micro-configurations onto a small set of stable macro-states $Z_t$. These include dimers, trimers, tetrahedra, octahedra, and 13-icosahedra. Such cluster forms have fewer degrees of freedom, show strong energetic stability, and occupy well-defined regions of state space.

On the macro level, the dynamics becomes an ε-machine. Causal states $E_i$ group past trajectories that yield identical future distributions: $P(Z_{t+1} \mid z(t|-\infty:0)) = P(Z_{t+1} \mid z'(t|-\infty:0))$. The transitions $T_{ij}=P(E_j|E_i)$ describe the full process. Because prebiotic cluster landscapes possess only a few deep attractors, the resulting ε-machine is small, near-deterministic, and exhibits characteristic attractor cycles and retrieval trajectories. ε-machines thus become natural tools for describing proto-informational dynamics: stable cluster geometries act as "symbol-like" states, and transitions encode the rules of an early prebiotic automaton. All three closure types integrate naturally into this layered model:

*Informational closure:* The future is predicted better from $Z_t$ than from $X_t$: $H(Z_{t+1} \mid Z_t) \ll H(Z_{t+1} \mid X_t)$.

*Causal closure:* Macro-level interventions produce the same transitions as micro-interventions that establish the same macro-state: $P(Z_{t+1} \mid do(Z_t)) = P(Z_{t+1} \mid do(X_t \in pre(Z_t)))$.

*Computational closure:* The ε-machine derived from the macroprocess has the same causal structure as the one obtained from the full micro-dynamics: $\varepsilon(f(X_t)) = \pi(\varepsilon(X_t))$.

Together, these elements form a unified four-level model of emergent prebiotic organization: physical forces act on the micro level; stable attractors arise on the meso level; these attractors form an ε-machine on the macro level; and attractor-coded memory with pattern retrieval yields a proto-informational "software layer" long before genes or enzymes existed.

## XI. CONCLUSION AND FUTURE WORK

Our analysis shows that Casimir–Lifshitz forces drive protocells into a small set of stable mesoscale attractors—dimers, tetrahedra, octahedra, and 13-icosahedra. These attractors persist despite microscopic noise and define a structured macro-dynamic. When coarse-grained, this dynamic becomes an ε-machine that is informationally and computationally closed in the sense of Rosas et al. [2]. The cluster geometries act as symbol-like states, and their transitions form a proto-algorithmic rule set.

This establishes a physically grounded route to the emergence of prebiotic information. The autonomous macro-dynamics of protocell clusters constitutes an early "software layer" encoded not in polymers but in attractor structure and pattern retrieval. In this view, information arises from the geometry, stability, and transition rules of the cluster state space.

Prebiotic protocell aggregates therefore offered a purely physical form of memory, function, and predictive structure—providing a substrate for later chemical and genetic evolution. Long before RNA or enzymes existed, mesoscale cluster dynamics already supported many of the informational properties required for biological complexity.